\documentstyle[twocolumn,epsf]{jpsj}

\newcommand{\I}{{\rm i}}
\newcommand{\EXP}{{\rm e}}
\newcommand{\nn}{\nonumber}

\title{Quantum Vortex in a Vectorial Bose-Einstein Condensate}
\author{
  Tomoya {\sc Isoshima},\thanks{E-mail: tomoya@mp.okayama-u.ac.jp}
  Kazushige {\sc Machida} and
  Tetsuo {\sc Ohmi$^{1}$}
}
\inst{Department of Physics, Okayama University, Okayama 700-8530, 

$^{1}$Department of Physics, Graduate School of Science, Kyoto University, Kyoto 606-8502}
\recdate{\today}
\abst{Quantum vortices in the multi-component 
Bose-Einstein condensation (BEC) are investigated theoretically.
It is found that three kinds of the  vortex configurations are possible
and their physical properties are discussed in details,
including the density distribution and the spin texture.
By using the Bogoliubov theory extended to the three component BEC,
the collective modes for these vortices are evaluated.
The local vortex stability for these vortices are examined in light of the 
existence of the negative eigenvalue, yielding a narrow magnetization window for
the local intrinsic stable region where the multi-components  work
together to stabilize a vortex  in a 
self-organized way. 
}
\kword{
Vortex,
Three component Bose-Einstein condensation,
Gross-Pitaevskii equation, Bogoliubov equation,
Collective modes
}

\begin{document}
\sloppy
\maketitle

\section{Introduction}

There have been much attention focused on Bose-Einstein condensation (BEC)
realized in alkali-metal atom gases.~\cite{firstRb,hulet,ketterle}
In particular, since two experimental groups~\cite{anderson1,anderson2,
madison1,madison2,madison3}
have succeeded in producing
quantum vortices in a rotating trapped BEC in $^{87}$Rb system,
many researchers investigate various 
aspects of this quantized vortex
in the hope to obtain the better understanding of
physical nature of superfluidity in this dilute weakly interacting Bose
system.
Here we have a better controlled mathematical tool, 
namely, Bogoliubov theory to treat it, compared with superfluid $^4$He 
where no established  microscopic theory to treat this strongly
interacting Bose condensed system is known.

In a series of our papers~\cite{isoshima1,isoshima2,isoshima3,isoshima4}
 we have been studying the problem
of the intrinsic instability of a vortex in BEC. 
We point out the existence of the negative eigenvalue in the Bogoliubov
excitation spectrum~\cite{isoshima1}, demonstrating that a
quantized vortex in BEC confined by a rigid wall is intrinsically
unstable and corresponding to the Rokhsar instability~\cite{rokhsar} due to
the spiraling-out of a vortex core in a harmonically trapped BEC.
Independently, the existence of the negative value
is pointed by Dodd {\it et al}.~\cite{dodd}
In order to arrest this instability and to stabilize it, we propose
several practical devices feasible experimentally, such as sending a laser
light at the core center to raise the potential which moves up
the whole Bogoliubov spectrum as a result, or raising temperature
and increasing the non-condensate fraction relative to the condensation which 
acts as an extra potential~\cite{isoshima3}.
A similar recovering method is proposed by Doi~\cite{doi-natsume}
who consider the two component BEC.
We also examine the external rotation effect~\cite{isoshima2} on the vortex stability and 
determine the several characteristic critical 
rotation frequencies~\cite{isoshima4},
including the lowest critical frequency which signals the first entry of a
vortex upon increasing the forced rotation.
This particular value~\cite{isoshima4} is by far nearer to the observed stirring frequency
in the recent vortex creation experiment by optical 
spoon~\cite{madison1,madison2,madison3}
than that estimated by the conventional Thomas-Fermi approximation.
This coincidence further encourages us to perform the present
study along this line.

So far in all the theories and experiments mentioned above the
internal degrees of freedom in an atom is frozen by the external magnetic 
field
serving as a confining potential.
By means of optical confinement, all the hyperfine states with F=1 in $^{23}$Na
active and three components with $m_{\rm F} = 1, 0, -1$ are
simultaneously Bose-condensed,~\cite{stenger}  giving rise to a new state of matter;
a BEC characterized by a vectorial order parameter.~\cite{stamperkurn}
Ohmi and Machida~\cite{ohmi_bec}, and Ho~\cite{ho} independently
introduce the basic Hamiltonian for describing this vector BEC by
extending the Bogoliubov framework and study the fundamental
properties of this interesting vector BEC, pointing out the richness of
the topological defect structures such as the $l$-vector textures,
domain wall structure, etc. These are very analogous to the $^3$He
superfluid A and B phases.~\cite{woelfle} 
Again as the basic interaction in these superfluid
Fermions is strong, we have to resort to
the strong coupling theory to describe it.
The existing theories for superfluid $^3$He
is quite phenomenological and difficult to derive microscopically.
In contrast, our theory here is quite microscopic and there is no room to
introduce phenomenological parameters in it.

Guided by this analogy, we are going to study here the
vortex structures in more details.
Yip\cite{yip_vortex} gave the first study on a vortex in the vectorial BEC,
classifying the possible vortex structures as functions
of the interaction strength of the spin channel and
the magnetization of the system.
Unfortunately, some of the vortices introduced by Yip (I and IV phase in
ref.\ \citen{yip_vortex}) are simply not allowed because the fundamental
phase requirement is violated for these particular
vortices (See eq.\ (\ref{eq:phase})).
Leonhardt and Volovik~\cite{aliceString} propose the
half-quantum vortex (Alice string).
Stoof~\cite{stoof} and Marzlin {\it et al}~\cite{marzlin} examine 
independently
the skyrmion creation in the vectorial BEC.
Thus, here continuing our study on the vectorial BEC\cite{ohmi_bec,
spindomain,isoshimanakahara1,isoshimanakahara2,double}, 
we perform
microscopic calculations of the possible vortices in harmonically trapped
vector BEC and examine the intrinsic stability problem in connection with
our intrinsic instability of a vortex in the scalar BEC.

After giving the Gross-Pitaevskii equation for the 
vectorial BEC with the three component order parameters in next Section,
we examine the fundamental phase requirements for the 
order parameter components when they form a vortex.
We enumerate several possible vortices in a cylindrical
symmetric case in \S 3.
In \S 4 each of these possible vortices is studied in details: 
their physical properties such as the density distributions of 
the three components 
and spin configuration
around a vortex core.
Evaluating the excitation spectrum for each vortex form 
by solving the Bogoliubov equation extended to the three components,
 we examine the stability condition
of each vortex as a function of the total
magnetization in \S 5. 
The last Section is devoted to summary and discussions.

\section{Gross-Pitaevskii Equation for Three Components}\label{sec:gp}

We treat a system of the Bose condensates with the internal
degrees of freedom $F=1$ so that the order parameter is characterized 
by the spins: $m_F=1,0,-1$.
We start with a general Hamiltonian
which fully considers system's symmetries
in real space and spin space:
\begin{eqnarray}
   H &=&
   \int\! {\rm d}{\bf r} [
      \sum_{i}\Psi_i^{\dagger}
          \left\{h({\bf r}) - \mu_{i}\right\}
      \Psi_i
\nn\\&&\  %
      + \frac{g_{\rm n}}{2} \sum_{ij}
         \Psi_i^{\dagger} \Psi_j^{\dagger} \Psi_j \Psi_i 
\nn\\&&\  %
      + \frac{g_{\rm s}}{2} \sum_{\alpha}
          \sum_{ijkl} \Psi_i^{\dagger}\Psi_j^{\dagger}
          (F_{\alpha})_{ik}(F_{\alpha})_{jl}
          \Psi_k \Psi_l
      ],
\label{eq:ham}
\end{eqnarray}
where 
\begin{eqnarray}
   h({\bf r}) &=& - \frac{\hbar^2 \nabla^2}{2m} + V({\bf r}),
\\
   g_{\rm n} &=& \frac{4 \pi \hbar^2}{m} \cdot \frac{a_0 + 2a_2}{3},
\\
   g_{\rm s} &=& \frac{4 \pi \hbar^2}{m} \cdot \frac{a_2 - a_0}{3} .
\end{eqnarray}
The subscripts are $\alpha = (x,y,z)$ and $i,j,k,l = (0, \pm1)$
corresponding to the above three species.
The chemical potentials for the three components;
$\mu_i$ $(i = 0, \pm1)$ 
satisfy $\mu_1 + \mu_{-1} = 2 \mu_{0}$.
The scalar field $V({\bf r})$ is the external confinement potential
such as an optical potential.

Following our previous paper,~\cite{ohmi_bec}
and the usual procedure for a BEC system without  spin,
the Gross-Pitaevskii equation extended to the three component
order parameters is obtained as
\begin{eqnarray}
   [
      \left\{
          h({\bf r}) - \mu_i + g_{\rm n} \sum_k |\phi_k|^2 
      \right\} \delta_{ij} 
&&\nn\\
      + g_{\rm s} \sum_{\alpha}\sum_{kl} \{
         (F_{\alpha})_{ij} (F_{\alpha})_{kl} \phi_k^{\ast} \phi_l
      \}
   ] \phi_j &=& 0.
\label{eq:gp}
\end{eqnarray}
These coupled equations for the $j$-th condensate
$\phi_j$ $(j = 0, \pm1)$ are used to calculate
various properties of a vortex in the following.

\section{Phase Requirements of Condensates}

The two-dimensional disk system is assumed
and the coordinate system  ${\bf r}=(r,\theta)$ is used.
The wavefunctions are divided into the radial part $\phi$ and
phase part $\gamma$ as
\begin{eqnarray}
   \phi_j(r, \theta) &=& \phi_j(r) \gamma_j(\theta) ,
\label{eq:phigamma1}
\\
   \gamma_j (\theta) &=& \exp[\I ( \alpha_j + \beta_j \theta ) ]
\label{eq:phigamma2}
\end{eqnarray}
where $j = 0, \pm 1$.
The wavefunctions $\phi_j(r)$ are real.
Temporarily we assume $\phi_j(r) \geq 0$.

The total energy of the system is given by
\begin{eqnarray}
E &=& \int {\rm d}^3{\bf r}
   [ \sum_j \phi_j(r) \frac{-\hbar^2}{2m}
      \left\{
          \frac{{\rm d}^2}{{\rm d}r^2} + \frac{1}{r}\frac{\rm d}{{\rm d}r}
          - \frac{( \beta_0 + j \beta )^2}{r^2}
      \right\} \phi_j(r)
\nn\\&&
\     + \sum_j\left\{
         V(r) + \frac{g_n}{2}\sum_k \phi_k^2(r)
         - (\mu_j - \mu_0)
      \right\} \phi_j^2(r)
\nn\\&&
\     + \frac{g_s}{2} \sum_\alpha \left
         \{\sum_{jk} \phi_j^\ast (F_\alpha)_{jk} \phi_k
      \right\}^2
   ].
\label{eq:E}
\end{eqnarray}
The phase coefficients $\gamma_j$ of the condensate wave function
are determined such that the
energy of the $g_s$ term in integrand of eq.\ (\ref{eq:E})
is minimized.
This is because the other terms in  eq.\ (\ref{eq:E}) are
not affected by the choice of phases.
The condition to minimize the $g_s$ term:
\begin{eqnarray}
   E_s
&=&
   \frac{g_s}{2} \sum_\alpha \left
         \{\sum_{jk} \phi_j^\ast (F_\alpha)_{jk} \phi_k
   \right\}^2
\nn\\
&=&
   \frac{g_s}{2} [
      2 \phi_{0}^2(r) \{
         \phi_{1}^2(r) + \phi_{-1}^2(r)
\nn\\&&
\ \      + \phi_1(r) \phi_{-1}(r) \left(
            \gamma_1 \gamma_{-1} \gamma_0^{\ast 2} +
            \gamma_{1}^\ast  \gamma_{-1}^\ast \gamma_0^{2}
         \right)
      \}
\nn\\&&
\  +  \left(  \phi_{-1}^2(r) - \phi_{1}^2(r)  \right)^2
   ]
  \label{eq:gsterm}
\end{eqnarray}
leads to
\begin{equation}
  \gamma_1 \gamma_{-1} \gamma_0^{\ast 2} = \pm1.
  \label{eq:phase}
\end{equation}
where the upper (lower) sign is used when $g_s < 0$ ($g_s > 0$).
This condition is rewritten in terms of $\alpha$ and $\beta$ as
\begin{eqnarray}
   2 \alpha_0 &=& \alpha_1 + \alpha_{-1} + n \pi
\\
   2 \beta_0 &=& \beta_1 + \beta_{-1} 
\end{eqnarray}
where $n$ is an integer.
We take $\alpha_0 = 0$ in the following.
Thus the phase factors of the wavefunctions are
generally written as 
\begin{equation}
   \left(
      \begin{array}{c} \gamma_1 \\ \gamma_0 \\ \gamma_{-1} \end{array}
   \right) =
   \left(
      \begin{array}{c}
         \exp[\I (\alpha + \beta \theta) ] \\
         1 \\
         \pm \exp[- \I (\alpha + \beta \theta) ] \\
      \end{array}
   \right)
   \exp[\I \beta_0 \theta]
\label{eq:gammas}
\end{equation}
with
\begin{equation}
   \beta = \beta_1 - \beta_0 = \beta_0 - \beta_{-1}
\label{eq:beta}
\end{equation}
and
\begin{equation}
  \alpha = \alpha_1 = -\alpha_{-1}.
\label{eq:alpha}
\end{equation}

In the followings we exclude the vortex states with winding number 
$\beta_i$ larger than 1 simply because the smaller winding number,
the more stable.
The possible combinations for
$\beta_i$ are
$(\beta_1, \beta_0, \beta_{-1})$ =
$(1, 0, -1), (1, 1, 1)$ and $(1, \frac{1}{2}, 0)$.
Because the value ``$\frac{1}{2}$'' does not allowed in this
cylindrically symmetric system,
this means that this component vanishes.
We write $(1, \frac{1}{2}, 0)$ as $(1, \mbox{none}, 0)$ in the following.
%
%
We call this vortex state the Alice state.
The Alice string discussed by Leonhardt and Volovik~\cite{aliceString} 
has this
combination of the winding numbers for the three components,
although the Alice string has another condition for the amplitudes.
The trivial combinations which are equivalent to the above ones, i.e.
$(-1, 0, 1), (-1, -1, -1)$, $(-1, \mbox{none}, 0)$ and
 $(0, \mbox{none}, \pm 1)$ are excluded.

The sign of $\gamma_{-1}$ in eq.\ (\ref{eq:gammas}) is determined
by the sign of $g_{\rm s}$
as shown in derivation of eq.\ (\ref{eq:gammas}).
When $g_{\rm s} < 0$ (ferromagnetic case), 
$\gamma_{-1} = + \exp[- \I (\alpha + \beta \theta) ]$.
When $g_{\rm s} > 0$ (antiferromagnetic case), 
$\gamma_{-1} = - \exp[- \I (\alpha + \beta \theta) ]$.
We redefine $\phi_{-1}$ and $\gamma_{-1}$
for simplicity of our calculation:
\begin{equation}
      \left\{ \begin{array}{@{}ll}
         \phi_{-1}(r) \geq  0  & (\mbox{ferromagnetic case})  \\
         \phi_{-1}(r) \leq  0  & (\mbox{antiferromagnetic case}) 
      \end{array} \right.  
\end{equation}

\begin{equation}
   \gamma_{-1} = \exp[- \I (\alpha + \beta \theta) ]
\label{eq:newgamma}
\end{equation}
The Gross-Pitaevskii equation eq.\ (\ref{eq:gp}) for the real
wavefunctions $\phi_j(r)$ becomes
\begin{eqnarray}
   [
   \frac{- \hbar^2}{2m}\left\{
      \frac{{\rm d}^2}{{\rm d}r^2}
      +\frac{1}{r}\frac{\rm d}{{\rm d}r}
      -\frac{(\beta_0 + j \beta)^2}{r^2}
   \right\}\delta_{ij}
&&\nn\\
   - \mu_i \delta_{ij}
   + g_n \sum_k \phi_k^2 \delta_{ij}
&&\nn\\
   + g_s \sum_{\alpha} \left\{
       (F_{\alpha})_{ij} \sum_{kl} (F_{\alpha})_{kl} \phi_k(r) \phi_l(r)
   \right\}
   ]
\phi_j(r)
&=& 0.
\label{eq:gp2}
\end{eqnarray}
The phase $\alpha$ does not appear here.
This means that $\phi_j(r)$ are independent of $\alpha$.

We have done the extensive calculations to solve the above equations.
To concentrate on a long straight vortex,
we assume that the system is uniform along the  $z$ axis as mentioned.
Since in the ferromagnetic case the phase separation occurs,
the results for a disk shape here may be appropriate to this case.

The properties of $F=1$ condensate is
classified into ferromagnetic and antiferomagnetic
depending on sign of $g_s$.
To represent antiferromagnetic ($ g_{\rm s}<0 $)
and ferromagnetic($ g_{\rm s}<0 $) case,
$g_{\rm s} = +0.1$ and $g_{\rm s} = -0.1$ are used as examples. 

The actual numerical computations are performed under
the conditions:
The mass and the scattering length of atoms are
$m = 3.81 \times 10^{-26} {\rm kg}$ and $a_0 = 2.75 {\rm nm}$.
The area density of particle is $2 \times 10^4 (\mu m)^{-1}$.
These are appropriate for those of Na atom.
The other scattering length $a_2$ is defined so that
$g_{\rm s}$ becomes $\pm 0.1g_n$.
The trapping frequency is $\omega / (2\pi) = 200$Hz.
The energy is scaled by the trap unit $\hbar \omega$.
The confining potential is given by $V({\bf r})={1\over 2}m\omega^2r^2$.

We calculate a set of the physical quantities
under given parameters such as $g_{\rm s}$,
the relative polarization $M/N$ (see below for the definition)
which is determined by adjusting the chemical potentials,
and the combination of winding number $\beta$'s.
We vary the relative polarization from 0 to 1.
Using symmetries, this covers all the possible value of $M/N$.
Both ferromagnetic ($ g_{\rm s}=-0.1 $) and
antiferromagnetic  ($ g_{\rm s}=0.1 $) cases are comparatively 
studied in the following.

\section{Three Kinds of Vortices}

We calculate the three kinds of the vortex states for
 various magnetizations.
The number of atoms and the magnetization 
are given by $N = \sum_{i=1,0,-1} N_i$ and $M = \sum_{i=1,0,-1}  i N_i$
where $N_i = \int |\phi_{i}({\bf r})|^2 d{\bf r}$ are
the number of atoms in the spin state $i$.
The range of the relative polarization  is $-1 \leq M/N \leq 1$.
As for the $(1,0,-1)$ and $(1,1,1)$  configurations,
it is sufficient to consider the range $0 \leq M/N \leq 1$.

It is convenient to describe the condensates
in terms of  the three wavefunctions
$\phi_{\alpha}\ (\alpha = x, y, z)$
where the spin quantization axis is taken along the $\alpha$ direction:
\begin{equation}
   \left(
      \begin{array}{c}
         \phi_x \\ \phi_y  \\ \phi_z
      \end{array}
   \right) = 
   \left(
      \begin{array}{ccc}
         \frac{- 1}{\sqrt{2}}   & 0 & \frac{1}{\sqrt{2}}   \\
         \frac{- \I}{\sqrt{2}}  & 0 & \frac{- \I}{\sqrt{2}} \\
         0       &    1   &    0 
      \end{array}
   \right)
   \left(
      \begin{array}{l}
         \phi_1 \\
         \phi_0 \\
         \phi_{-1}
      \end{array}
   \right).
\end{equation}
Let define the  $m$- and $n$-vectors as
\begin{eqnarray}
   {\bf m} &=
   & (m_x, m_y, m_z)  = {\rm Re}( \phi_x, \phi_y, \phi_z ) \\
   {\bf n} &=& (n_x, n_y, n_z)  = {\rm Im}( \phi_x, \phi_y, \phi_z ).
\end{eqnarray}
where ${\bf m}$ and ${\bf n}$ are real vectors.
The $l$-vector which points the direction of the local magnetization
is ${\bf l} = {\bf m} \times {\bf n}$.
The corresponding unit vector is $\hat{\bf l}$.
%
%

\subsection{Alice states}

When the phase configuration of the condensates is the Alice type:
\begin{equation}
\left(
      \begin{array}{c}
         \phi_1 \\ \phi_0 \\ \phi_{-1}
      \end{array}
\right) = 
\left(
      \begin{array}{l}
         \phi_1(r) \EXP^{\I \theta} \\ 0  \\ \phi_{-1}(r)
      \end{array}
\right),
\end{equation}
the $l$-vector is always parallel to the $z$-axis:
$l_x = l_y = 0$, $l_z = \phi_1^2 - \phi_{-1}^2$. 
The zero component of the condensate $\phi_0$ is always zero
and the system is the 2-component, consisting of $\phi_1$ and $ \phi_{-1}$.

\begin{figure}
   \begin{center}\leavevmode
   \epsfxsize=6cm \ \epsfbox{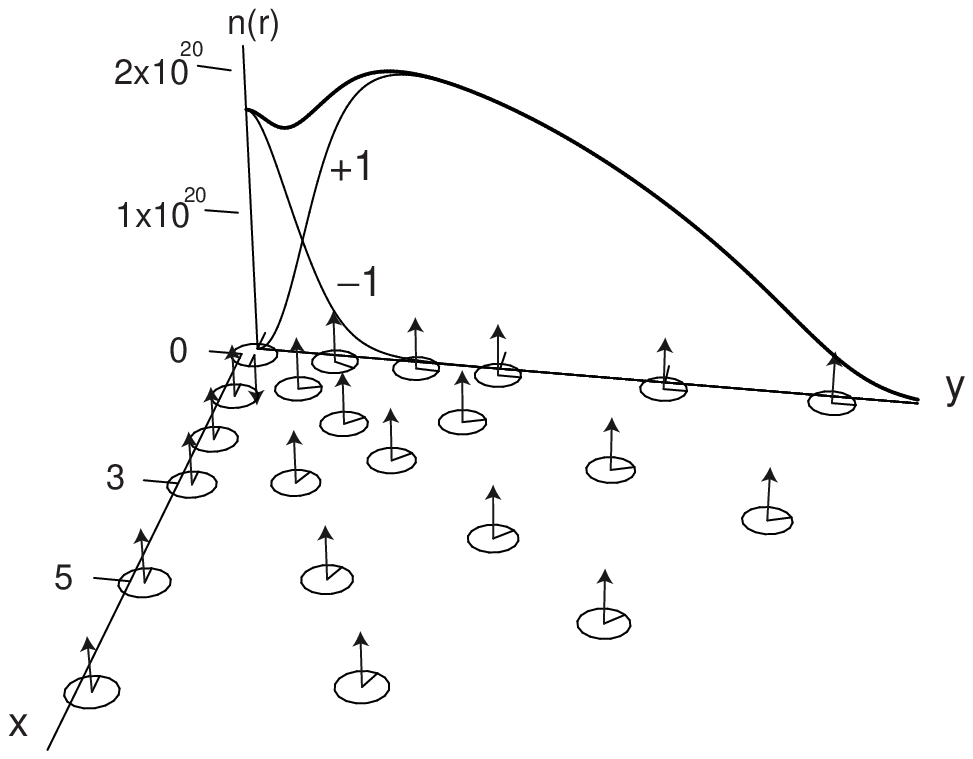}\\
   (a)\\
   \epsfxsize=6cm \ \epsfbox{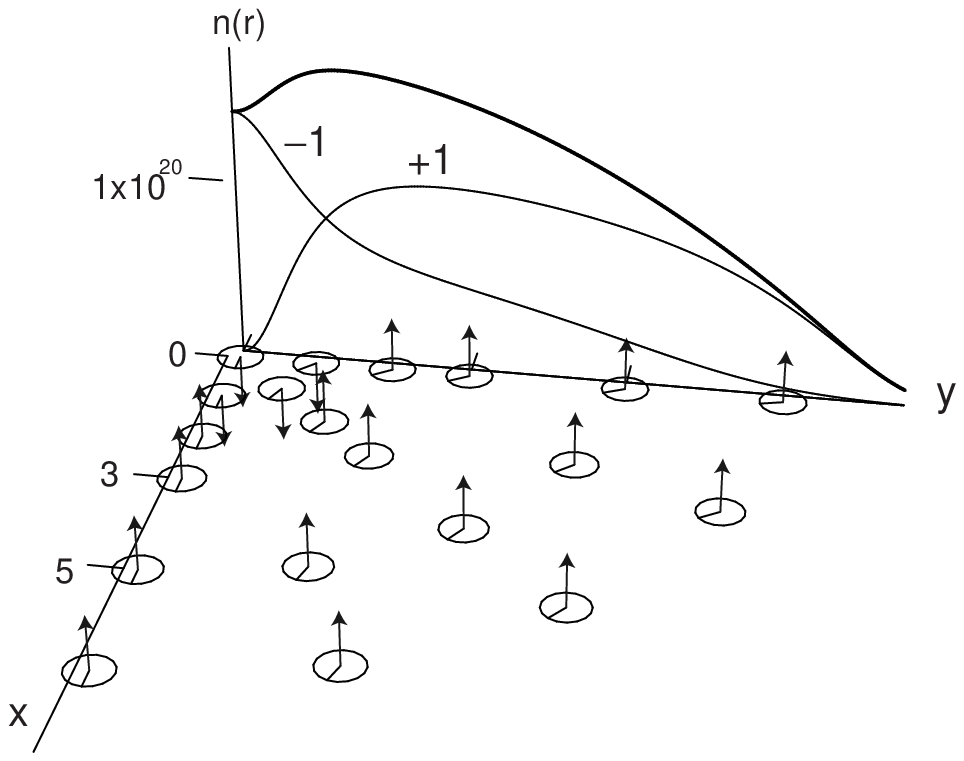}\\
   (b)\\
  \end{center}
   \caption{ 
The density profiles of the  condensates are displayed along the
radial direction and the $\hat{l}$-vector
in the Alice type vortex is shown for the quadrant in the ($x,y$) plane.
The vertical axis is the density of the  condensates.
The upper bold curve shows the total density $n(r)$
and the thin curves show the density of each component $n_i(r)$.
The arrows on the ($x,y$) plane are the $\hat{l}$-vectors.
The small circles are perpendicular to each $\hat{l}$-vector
and the short rods show $m$-vectors.
They are depicted at $r=\sqrt{x^2 + y^2} = 0,1,2,3,5,\mbox{ and } 7\ \mu \rm{m}$.
(a) Ferromagnetic case. The magnetization $M/N= 0.967$.
(b) Antiferromagnetic  case. $M/N = 0.531$.
Note that the intermediate magnetization like this is not possible
for the ferromagnetic case.
   }
\label{fig:alice}
\end{figure}

When the interaction is ferromagnetic,
it is conceivable that one of the two components vanishes
and the condensate reduces to the single component system.
The system hardly has intermediate magnetization.
However, when the +1 component carries the phase winding,
the other component may appear at the center of the vortex.
In fact, as shown in Fig.\ \ref{fig:alice}(a) where the density profiles
of the Alice state are displayed, the core region of the +1 component 
is filled by the remaining component -1.
The range of magnetization is $M/N = -1, \mbox{ and from } +0.967 \mbox{ to } +1$.
In the intermediate range $-1 < M/N < +0.967$,
the $\phi_0$ component, which is not allowed in this
cylindrically symmetric system, is likely to appear due to
spin interaction term eq.~(\ref{eq:gsterm})
and cylindrical symmetry solution of
GP equation does not obtained.~\cite{martikainen}
%

When the interaction is antiferromagnetic,
the amplitude ratio of the two components may vary continuously,
depending on the magnetization.
Figure\ \ref{fig:alice}(b) shows the case when $M/N = 0.531$.
For  all ranges of $M/N$, 
the density of the +1 component has a  peak at $r \simeq  3 \mu m$.

\subsection{(1, 0, -1) vortex}

\begin{figure}
   \begin{center}\leavevmode
   \epsfxsize=6cm \ \epsfbox{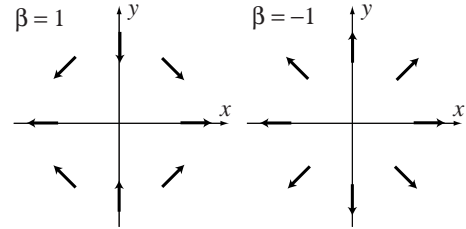}\\
   \end{center}
   \caption{ 
   The bold arrows show x and y component of
   $l$-vectors when winding number combination is $(1,0,-1)$
   for the ferromagnetic case ($\alpha = 0$).
   The $l$-vectors show cross disgyration when $\beta = 1$
   and radial disgyration when $\beta = -1$.
   For the antiferromagnetic case the direction of the  $l$-vectors
   in the $(x,y)$ plane is reversed.
   }
\label{fig:lvec:101state}
\end{figure}

In this vortex, the condensate is described by
\begin{equation}
\left(
    \begin{array}{c}
        \phi_1 \\ \phi_0 \\ \phi_{-1}
    \end{array}
\right) = 
\left(
    \begin{array}{l}
        \phi_1(r)    \EXP^{  \I(\beta\theta + \alpha) } \\
        \phi_0(r)  \\
        \phi_{-1}(r) \EXP^{- \I(\beta\theta + \alpha) }
    \end{array}
\right).
\end{equation}
The winding number of $\phi_0$ is zero and
the 0 component is only the component of the condensates
which  can exist at the vortex center.
The magnetization of the  condensate can vary continuously in this
configuration of the winding number.
The $l$-vector is given by
\begin{eqnarray}
\left(
    \begin{array}{c}
       l_x \\ l_y
    \end{array}
\right) &=&
\phi_0(r) \frac{\phi_1(r) + \phi_{-1}(r)}{\sqrt{2}}
\left(
    \begin{array}{c}
       \cos (\alpha + \beta \theta) \\
       - \sin(\alpha + \beta \theta) \\
    \end{array}
\right)
\\
    l_z &=& \frac{\phi_1^2(r) - \phi_{-1}^2(r)}{2}
\end{eqnarray}
Figure\ \ref{fig:lvec:101state} shows the $l_x$ and $l_y$ configurations in
the $(x,y)$ plane for the two kinds of the disgyration with either $\beta=1$
or $\beta=-1$.

\begin{figure}
   \begin{center}\leavevmode
   \epsfxsize=6cm \ \epsfbox{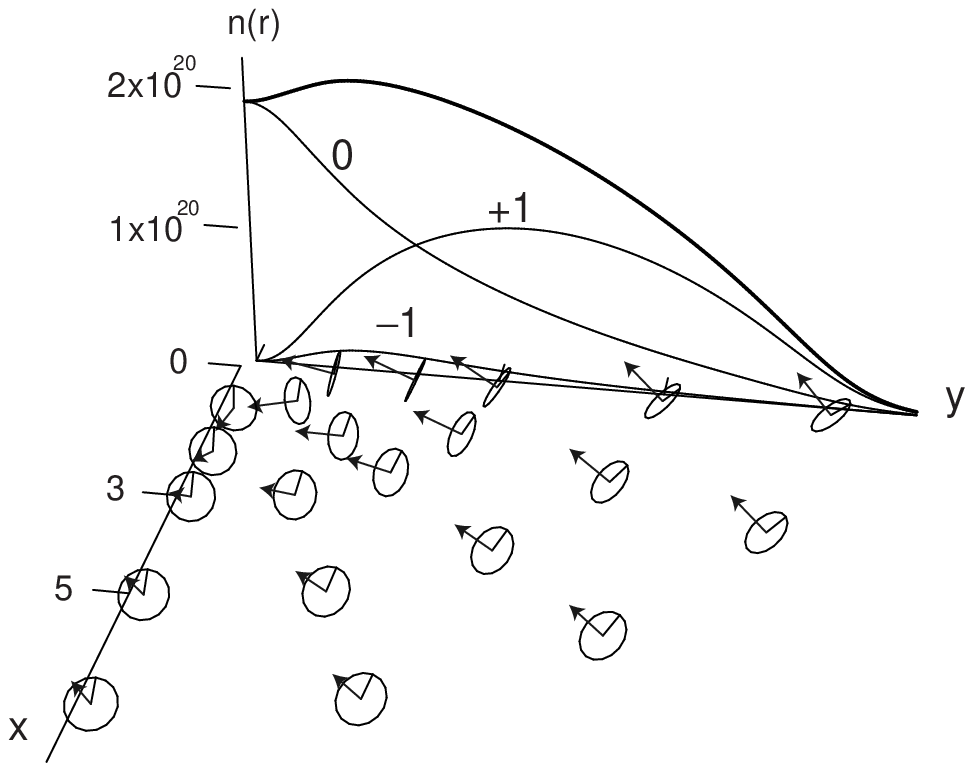}\\
   (a)\\
   \epsfxsize=6cm \ \epsfbox{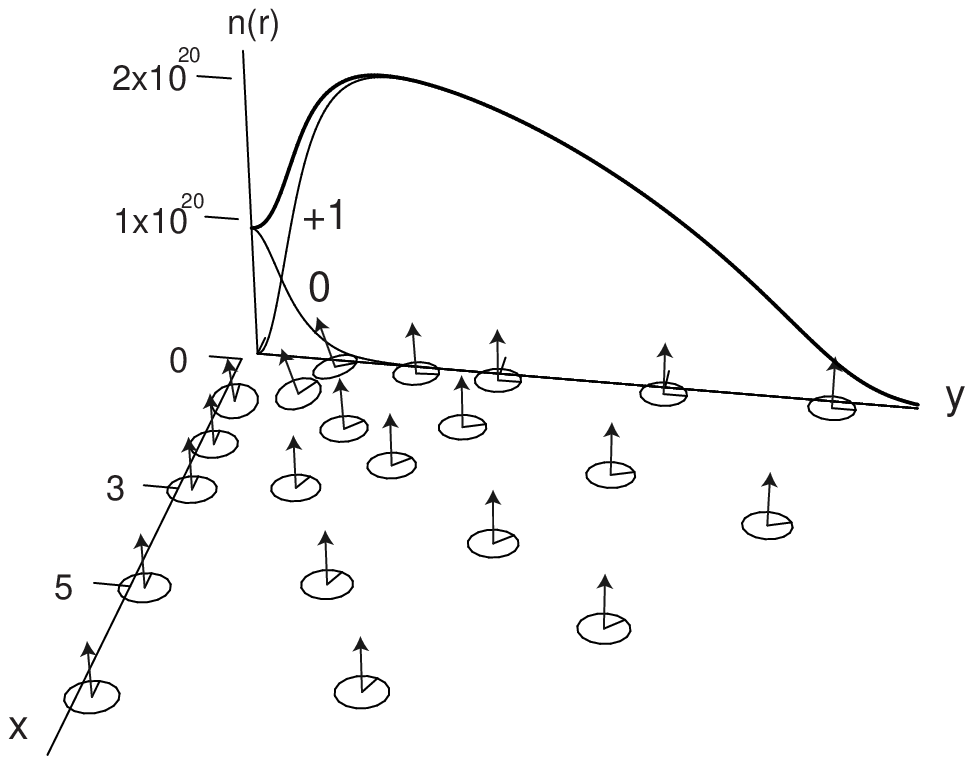}\\
   (b)\\
  \end{center}
   \caption{ 
The density profiles of the condensates and the $\hat{l}$-vector
when the interaction is ferromagnetic and
the winding number combination is $(1,0,-1)$.
The vertical axis is the density of the condensate.
The upper bold curve shows the total density $n(r)$
and the thin curves show the density of each component $n_i(r)$.
(a) $M/N = 0.612$.
(b) $M/N = 0.994$.
   }
\label{fig:101state:ferro}
\end{figure}

Figures\ \ref{fig:101state:ferro}(a) and (b)
show the particle density profiles along the radial direction
in the ferromagnetic case. 
It is seen from Fig.\ \ref{fig:101state:ferro}(a) 
that the $\phi_0$ component fills in the core region, 
thus there is no appreciable particle number depleted region near the center.
Notice  that $\phi_{-1}(r) \leq  0$ when the interaction is ferromagnetic.

\begin{figure}
   \begin{center}\leavevmode
   \epsfxsize=6cm \ \epsfbox{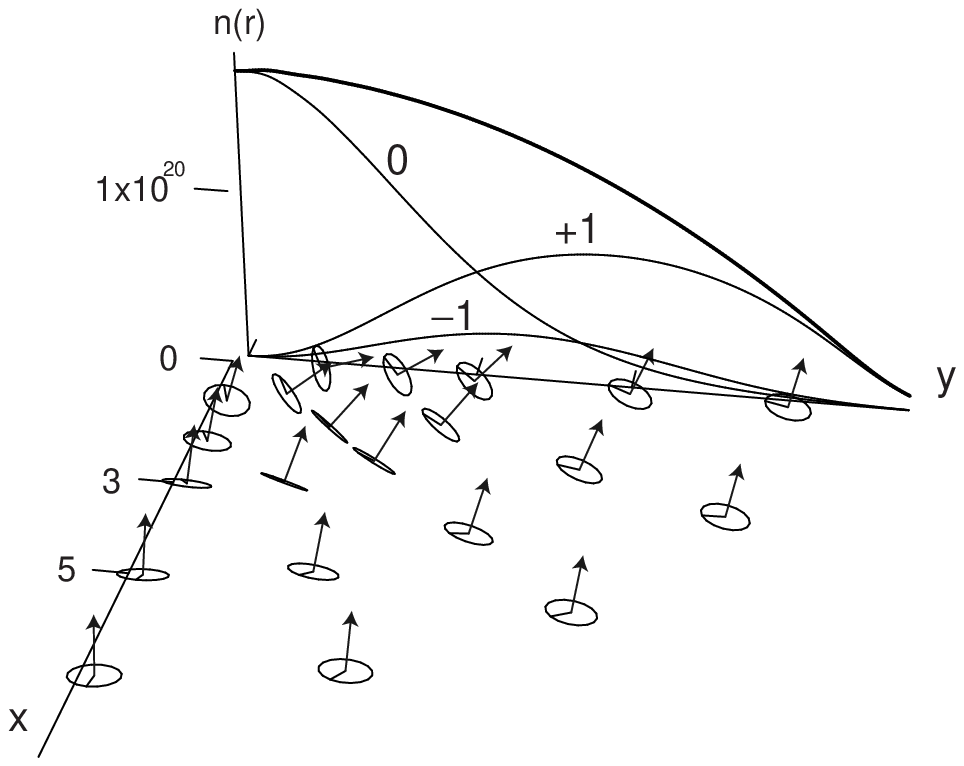}\\
   (a)\\
   \epsfxsize=6cm \ \epsfbox{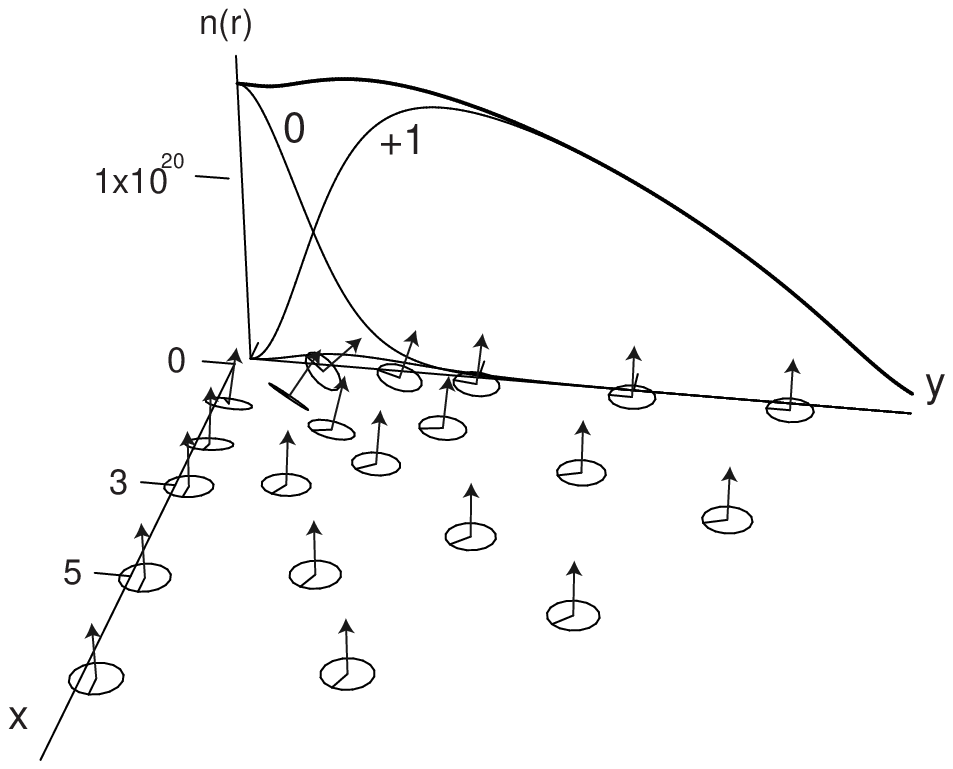}\\
   (b)\\
  \end{center}
   \caption{ 
The density profiles of the condensates and the $\hat{l}$-vector
when the interaction is antiferromagnetic and
the winding number combination is $(1,0,-1)$.
The vertical axis is the density of the condensate.
The upper bold curve shows the total density $n(r)$
and the thin curves show the density of each component $n_i(r)$.
(a) $M/N = 0.474$.
(b) $M/N = 0.952$.
   }
\label{fig:101state:anti}
\end{figure}

The corresponding antiferromagnetic case for the (1,0,-1)
vortex is displayed in
Fig.\ \ref{fig:101state:anti}(a) and (b).
Here it is also true that
the large number of the $\phi_0$ component occupies the core region,
pushing out $\phi_1$ and $\phi_{-1}$. The total particle number almost
smoothly decreases outwardly.
%

\subsection{(1,1,1) vortex}

\begin{figure}
   \begin{center}\leavevmode
   \epsfxsize=6cm \ \epsfbox{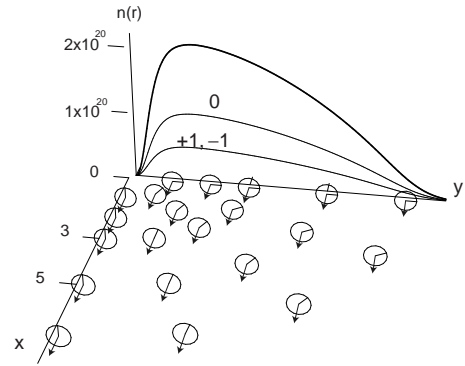}\\
  \end{center}
   \caption{ 
The density profile of the condensate and the $\hat{l}$-vector
for the  ferromagnetic case of the $(1,1,1)$ vortex.
The magnetization $M/N = 0$.
The vertical axis is the density of the condensate.
The upper bold curve shows the total density $n(r)$
and the thin curves show the density of each component $n_i(r)$.
   }
\label{fig:111state}
\end{figure}

All of components have the winding number 1 in
this vortex, namely, the wavefunction is written in the form: 
\begin{equation}
\left(
      \begin{array}{@{}c@{}}
         \phi_1 \\ \phi_0  \\ \phi_{-1}
      \end{array}
\right) = 
\left(
      \begin{array}{@{}l@{}}
         \phi_1(r)  \\ \phi_0(r)  \\ \phi_{-1}(r)
      \end{array}
\right)\EXP^{\I \theta}
\end{equation}
It is shown that the $l$-vector does not depend on $\theta$.
\begin{equation}
   {\bf l} =
   \left( \begin{array}{c}
      \phi_0 \frac{\phi_1 + \phi_{-1}}{\sqrt{2}} \cos (\alpha) \\
      \phi_0 \frac{\phi_1 + \phi_{-1}}{\sqrt{2}} \sin (- \alpha) \\
      \frac{1}{2} ( \phi_1^2 - \phi_{-1}^2 )
   \end{array}\right)
\label{eq:111:lvec}.
\end{equation}
For the ferromagnetic case, this reduces to single component BEC. 
Figure\ \ref{fig:111state} shows the typical results for this case,
namely, as all the components have non-vanishing phase winding, the densities
vanish at the vortex core center. The total density distribution 
looks like a vortex in the
scalar BEC.
For the antiferromagnetic case, 
the $l$-vector given by  eq.\ (\ref{eq:111:lvec}) vanishes when
$|\phi_{-1}| = |\phi_{1}|$.

\section{Local Stability of Vortex State}\label{sec:popov}

\subsection{Formulation}

In order to determine  the intrinsic stability condition
for each vortex state without the external rotation, 
we give the formulation for evaluating the excitation spectrum of 
the three component BEC, which is based on the Bogoliubov theory
extended to take into account the spin degrees of freedom.
The collective excitation spectrum of the condensate, whose static properties are
determined  by the Gross-Pitaevskii equation eq.~(\ref{eq:gp}),
is a solution of the Bogoliubov equations given by
\begin{eqnarray}
   \sum_j \{
        A_{ij}     u_q({\bf r}, j)
      - B_{ij}     v_q({\bf r}, j)\} &=& \varepsilon_q u_q({\bf r}, i),
\label{eq:bogo1}
\end{eqnarray}
\begin{eqnarray}
   \sum_j \{
        B^{\ast}_{ij}   u_q({\bf r}, j)
      - A^{\ast}_{ij}   v_q({\bf r}, j)\} &=& \varepsilon_q v_q({\bf r}, i)
\label{eq:bogo2}
\end{eqnarray}
where
\begin{eqnarray}
   A_{ij} &=&
      h({\bf r}) \delta_{ij} - \mu_i \delta_{ij}
\nn\\&&
      + g_n \left\{
         \sum_k  |\phi_k|^2  \delta_{ij}
         + \phi_i \phi_j^{\ast}
      \right\}
\nn\\&&
      + g_{\rm s}
         \sum_{\alpha}\sum_{kl} [
            (F_{\alpha})_{ij} (F_{\alpha})_{kl} \phi_k^{\ast} \phi_l
\nn\\&&\quad
            +(F_{\alpha})_{il} (F_{\alpha})_{kj} \left(
               \phi_k^{\ast} \phi_l
            \right)
         ],
\label{eq:a}
\\
   B_{ij} &=&
      g_{\rm n} \phi_i \phi_j
      +g_{\rm s} 
         \sum_{\alpha} \sum_{kl} \left[
            (F_{\alpha})_{ik} \phi_k  (F_{\alpha})_{jl} \phi_l
         \right],
\label{eq:b}
\end{eqnarray}
$u_q({\bf r}, i)$ and $v_q({\bf r}, i)$ are
the $q$-th eigenfunctions with the spin $i$ and
$\varepsilon_q$ corresponds to the $q$-th eigenvalue.

We are treating a cylindrical system ${\bf r} = (r, \theta, z)$.
The relevant physical quantities are written in the
following form up to the phase factor:
\begin{eqnarray}
    u_q({\bf r}, j)    &=&
    u_q(r, j)
      \EXP^{\I (q_{\theta} + \beta_0) \theta}
      \EXP^{\I (\alpha +  \beta \theta)j },
\\
    v_q({\bf r}, j)    &=&
    v_q(r, j)
      \EXP^{\I (q_{\theta} - \beta_0) \theta}
      \EXP^{- \I (\alpha +  \beta \theta)j },
\end{eqnarray}
%
with the angular momentum  $q_{\theta}$. 
(The momentum $q_z$ along the $z$ direction is ignored.)
%
%

\subsection{Stable range}

\begin{figure}
  \begin{center}\leavevmode
    \hspace{-0.5cm}
    \epsfxsize=4.3cm \ \epsfbox{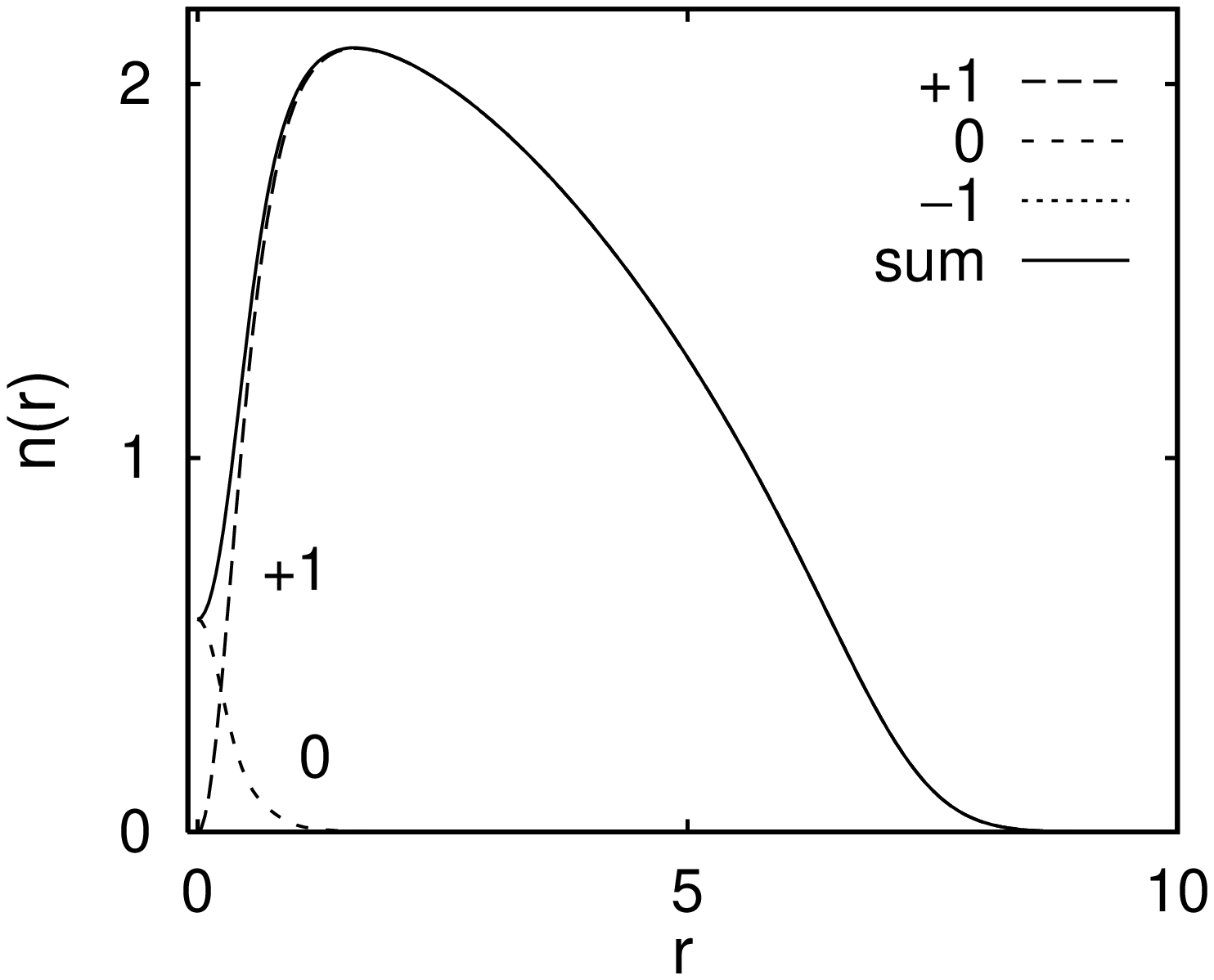}\hspace{-0.5cm}
    \epsfxsize=4.3cm \ \epsfbox{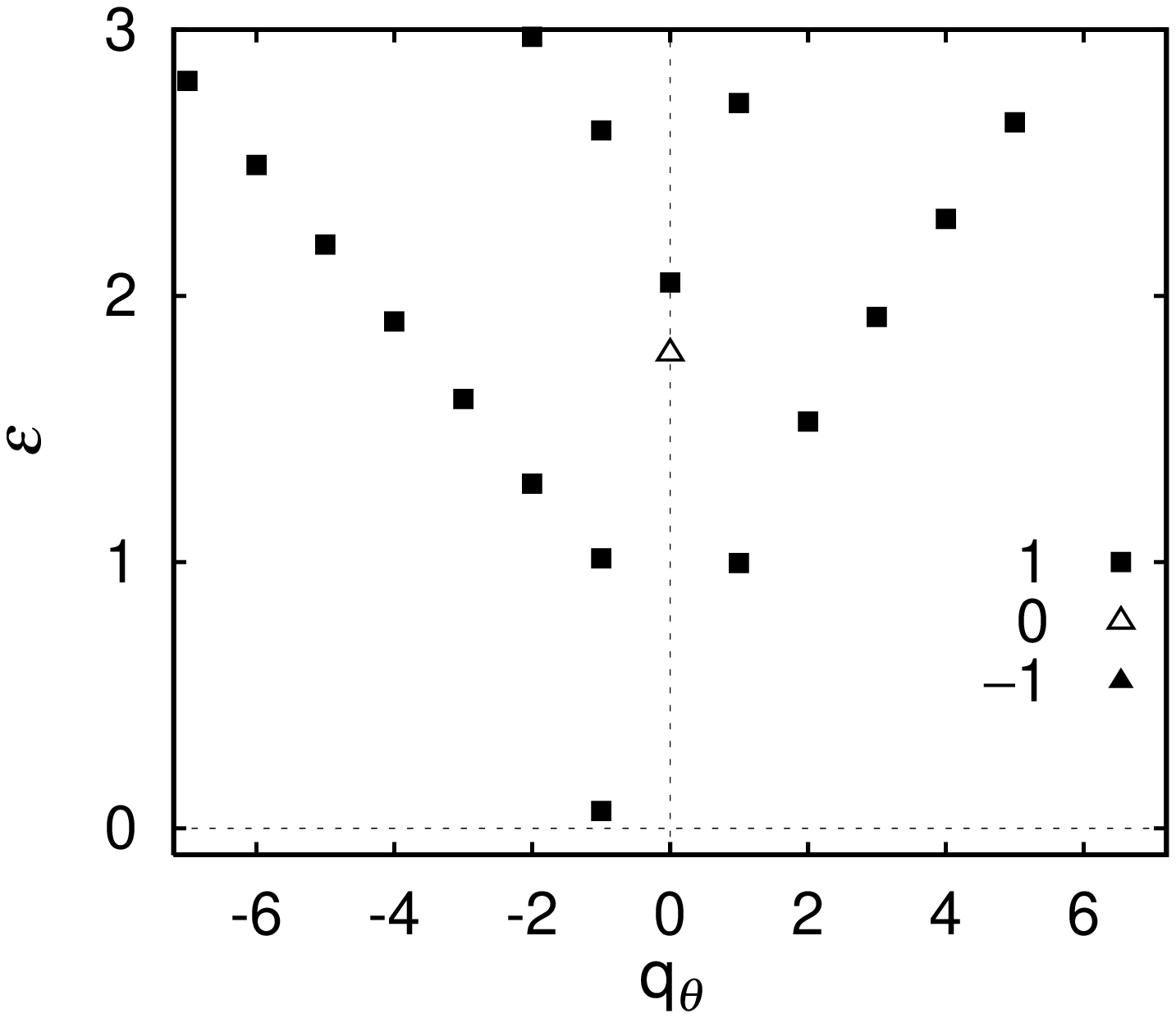}\\
    (a)\hspace{3.5cm}(b)
  \end{center}
  \vspace{0.5\baselineskip}
  \begin{center}\leavevmode
    \hspace{-0.5cm}
    \epsfxsize=4.3cm \ \epsfbox{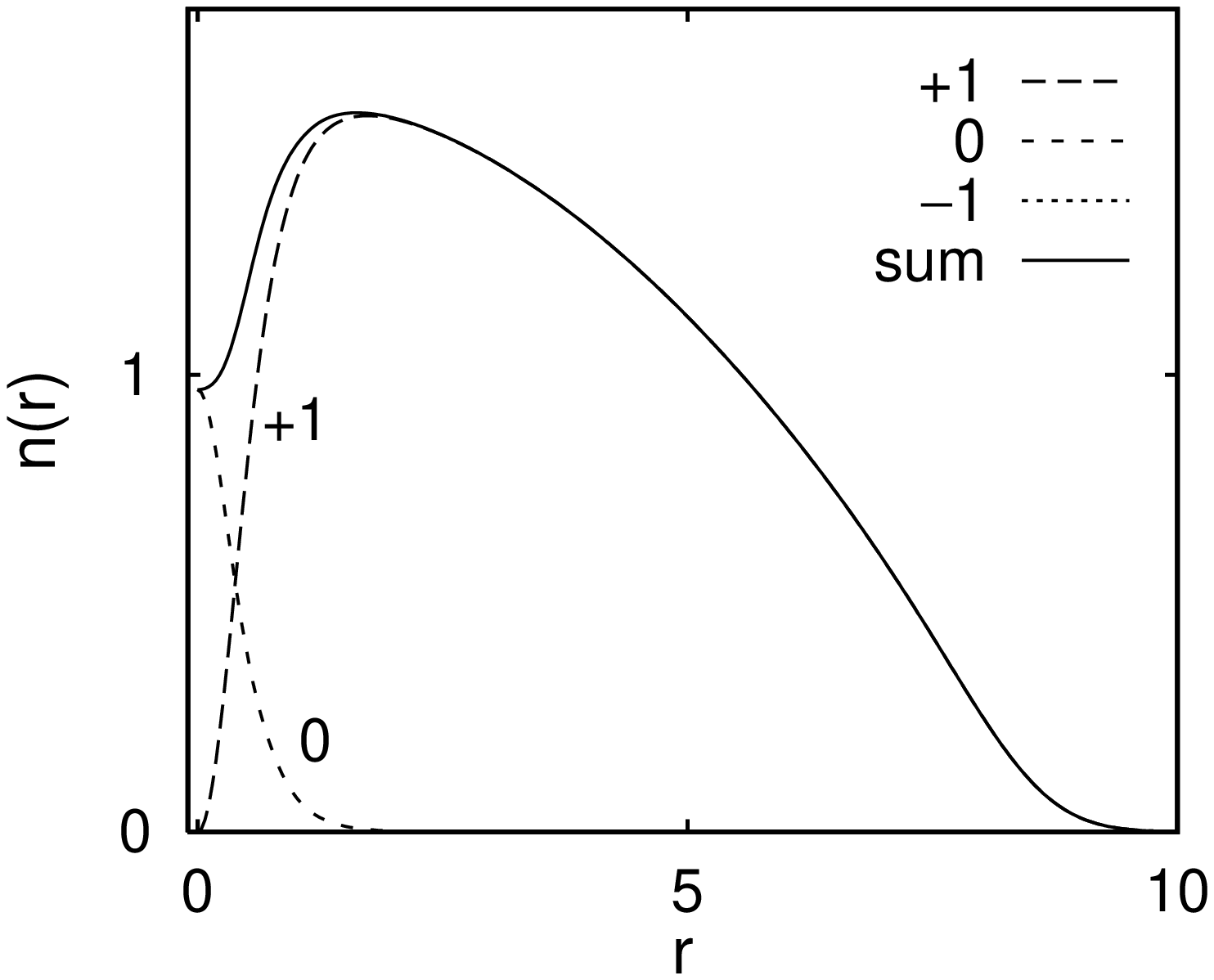}\hspace{-0.5cm}
    \epsfxsize=4.3cm \ \epsfbox{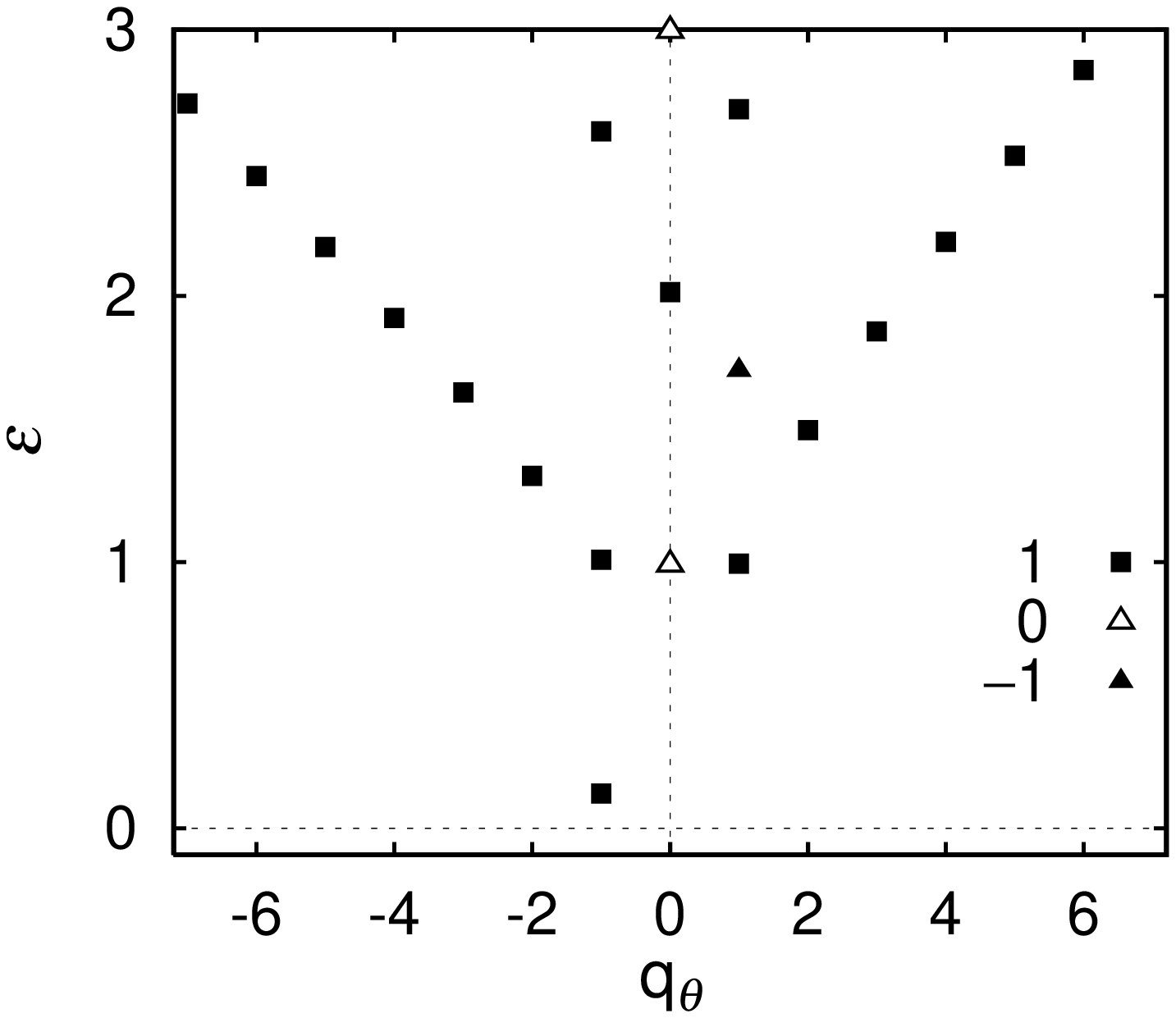}\\
    (c)\hspace{3.5cm}(d)
  \end{center}
  \vspace{0.5\baselineskip}
  \begin{center}\leavevmode
    \hspace{-0.5cm}
    \epsfxsize=4.3cm \ \epsfbox{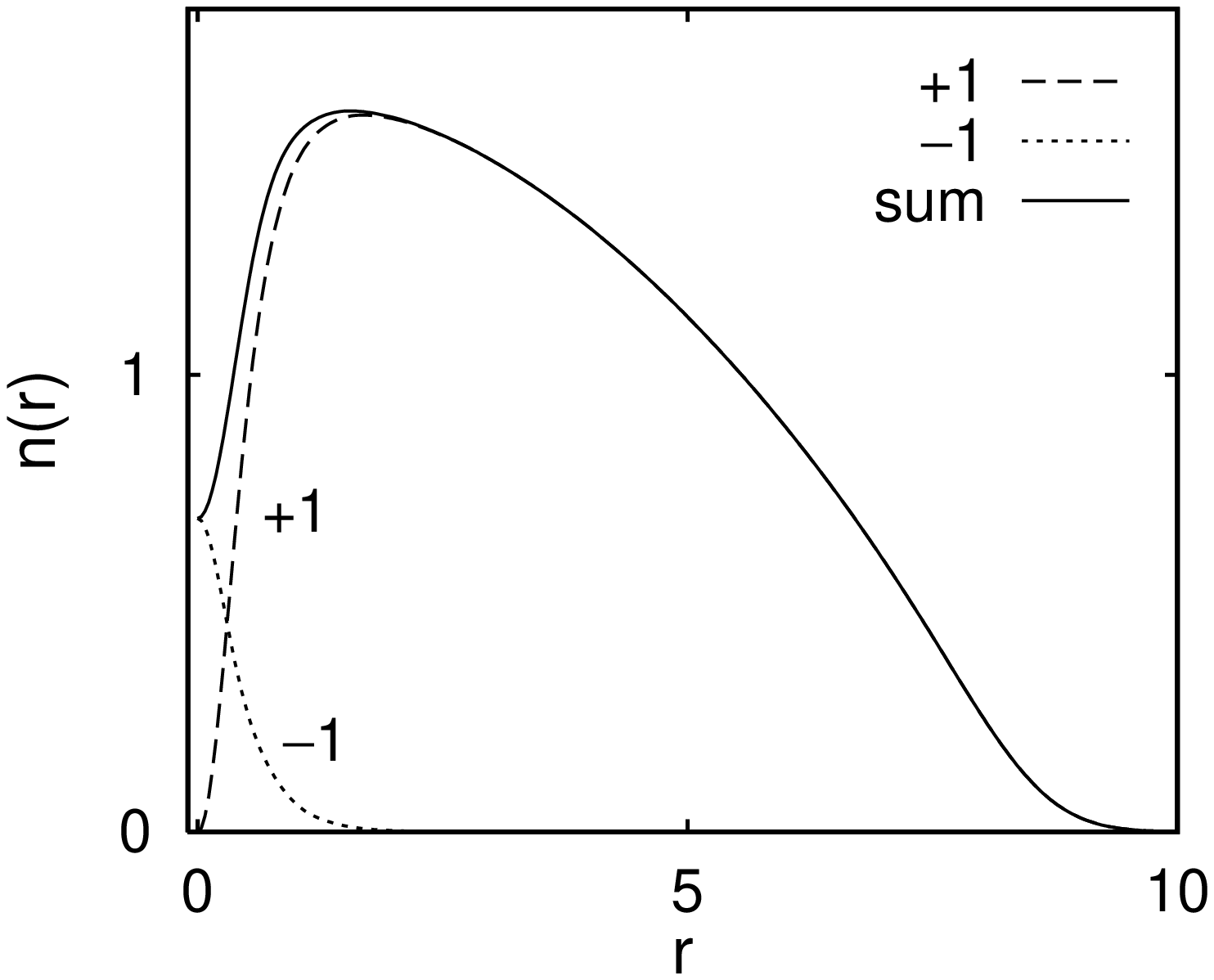}\hspace{-0.5cm}
    \epsfxsize=4.3cm \ \epsfbox{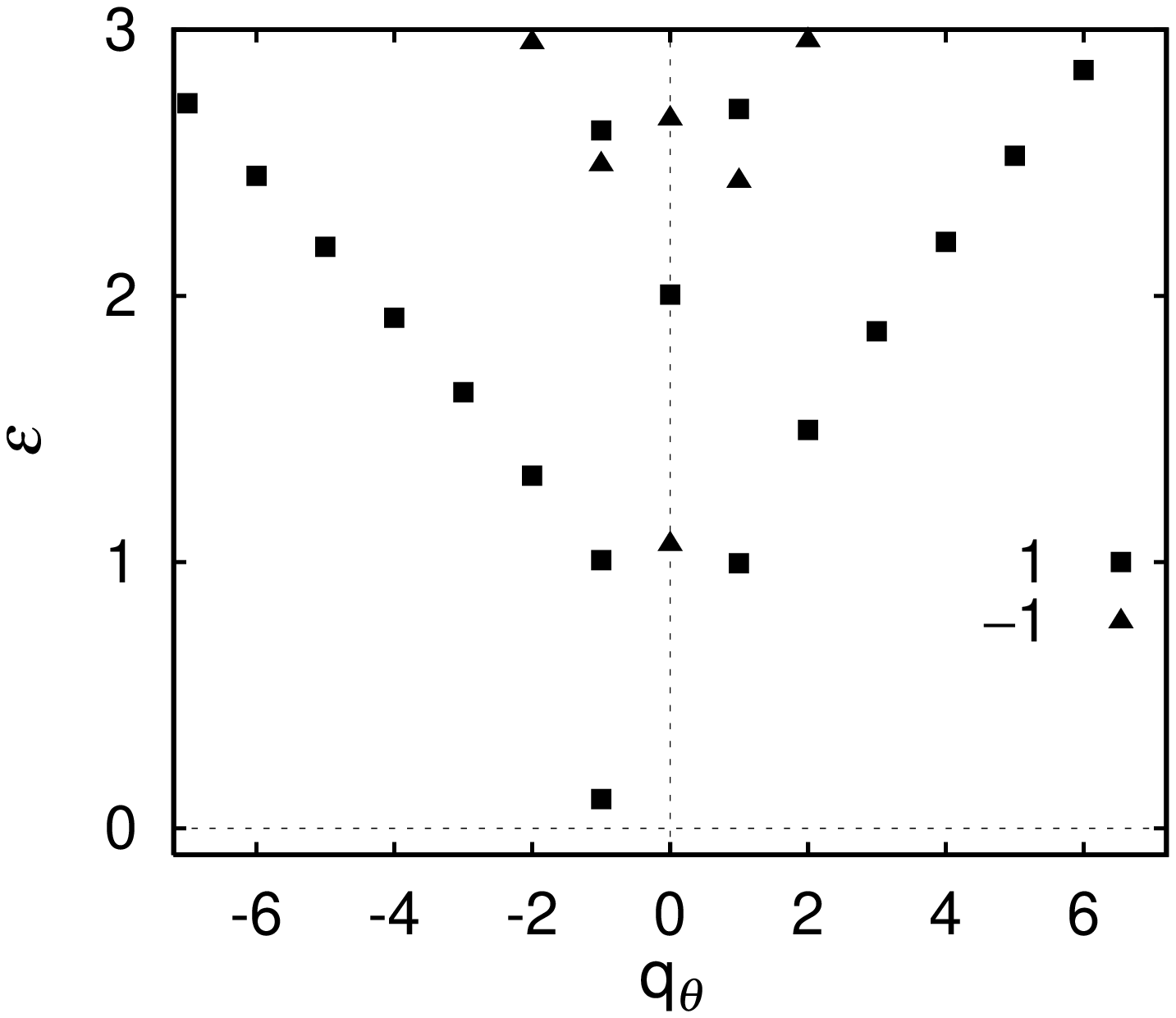}\\
    (e)\hspace{3.5cm}(f)
  \end{center}
  \vspace{0.5\baselineskip}
  \caption{
      The density profiles (in the units of $10^{20} {\rm m}^{-3}$ and $\mu m$);
      (a), (c) and (e),
      and the corresponding excitation spectra; (b), (d) and (f)
      for the locally stable vortices.
      The eigenvalues are normalized by $\hbar\omega$
      and characterized by different  marks,
      indicating where the eigenvalues belong.
      (a) and (b) Ferromagnetic (1,0,-1) vortex, $M/N = 0.998$.
      (c) and (d) Antiferromagnetic (1,0,-1) vortex, $M/N = 0.994$.
      (e) and (f) Antiferromagnetic Alice vortex, $M/N=0.991$.
   }
\label{fig:stable}
\end{figure}

When all of the excitation levels have the positive eigenvalue,
the vortex state is stable~\cite{isoshima1,isoshima2}, namely,
the vortex is locally stable in the energy configuration space. 
Thus the stable region of the magnetization 
for a given vortex is determined by evaluating 
the whole excitation spectrum of  the above Bogoliubov 
equations: eqs.~(\ref{eq:bogo1}) and ~(\ref{eq:bogo2}).
We calculate them to check if there exists the
negative eigenvalue. This procedure
yields the stable magnetization region.
It should be noted that  the lowest level at the particular 
angular momentum $q_{\theta} = -1$
in the excitation spectrum 
often becomes negative, signaling the local instability
of a  vortex
as discussed in the single component BEC.~\cite{isoshima1,isoshima4}
The positive spectrum does not
mean that this vortex is stable globally relative to other states, such as
non-vortex state. Let us discuss here the local stability of our enumerated vortices.
The stable ranges for the magnetization  where the negative mode does not appear
and system is stable for each vortex are listed below:

(1) For  the (1,0,-1)  vortex configuration in the 
ferromagnetic case,
the stable range of $M/N$ is determined as 
\begin{equation}
    0.992 < M/N < 0.999 \ \ (\mbox{ferromagnetic (1,0,-1) vortex}) 
\end{equation}
The typical density profile and excitation spectrum are shown in
Figs.\ \ref{fig:stable}(a) and \ref{fig:stable}(b) where the left hand side column
depicts the density distributions for each component and the 
right hand side column shows the corresponding excitation spectrum as a function
of  the angular momentum $q_{\theta}$.
The magnetization is $M/N = 0.998$ here.

(2) For the (1,0,-1) vortex in the 
antiferromagnetic case,
the stable range is given by
\begin{equation}
  0.988 < M/N < 0.999 \ \ (\mbox{antiferromagnetic (1,0,-1) vortex }) 
\end{equation}
We show the density profile and the excitation 
spectrum in Figs.\ \ref{fig:stable}(c) and \ref{fig:stable}(d).
The magnetization is $M/N = 0.994$ here.
Comparing the above ferromagnetic case,
because the magnetization decrease
the 0-component fills in more the vortex core region.
It pushes up the lowest excitation and stabilizes this vortex as a result.

(3) In the Alice vortex of the antiferromagnetic case,
the stable range is determined as
\begin{equation}
  0.972 < M/N < 0.999 \ \ (\mbox{antiferromagnetic Alice vortex }) 
\end{equation}
We display the density profile and the excitation spectrum for this
stable Alice vortex in Figs.\ \ref{fig:stable}(e) and \ref{fig:stable}(f).
The density distribution of the antiferromagnetic Alice vortex
is very similar to that in the (1,0,-1) vortex shown in Fig.\ref{fig:stable} (a) and (c).

It is rather remarkable that these multiple components work together 
so as to stabilize the vortex locally.
As the magnetization decreases from $M/N=1$,
the lowest eigenvalue at $q_{\theta} = -1$ becomes
positive.
Upon further decreasing beyond the critical magnetization,
its eigenvalue becomes negative again, indicating the local instability.
This was not seen in the single 
component case where the vortex is always locally unstable~\cite{isoshima1,isoshima4}.
This becomes stable only  when the system is subjected under external
rotation or when introducing the pinning potential.

It is generally seen throughout these vortices
that the component with the winding number 0 
fills in the core region of the 
+1 component with the winding 1.
This 0-winding component stabilize the vortex state.
For the two component BEC,
a similar stabilization mechanism is considered by Doi~\cite{doi-natsume}.

The locally stable regions are limited to
the cases where the magnetizations are close to the full polarized state.
If the magnetization increases further towards the complete polarization,
the system effectively becomes
one component BEC and shows the local instability
which was studied previously in details.~\cite{isoshima1,isoshima4}
This instability means the finite life time of the vortex state
which is observed experimentally.~\cite{anderson1,anderson2,madison1,madison2,madison3}

%

\section{Summary and Conclusion}

We have established a general framework to
calculate various physical properties of a vortex
in a cylindrical symmetric situation for a three component BEC 
system with $ m_F=1,0,-1$ of  $F=1$. This allows us 
to evaluate the density distributions of the multiple component
BEC, and the spatial variations of the triad $(m,n,l)$
which fully characterize a vortex.
This theoretical framework yields not only the above static properties,
but also the dynamical properties, such as the collective modes. 
The latter is used to examine the local stability of a vortex.

In the present paper, we have taken up three typical vortices which
might be realized in the actual experimental situations, 
namely, the phase windings of the wave functions for the three 
components $\phi_1, \phi_0$ and $\phi_{-1}$ are given by  (1,0,-1), (1,1,1)
and (1, none, 0) as the representative vortices.

We have characterized these vortices for both
ferromagnetic and antiferromagnetic interaction cases.
In general, the 0-winding component, if any, fills in 
the core region.  The densities of each component vary smoothly in 
the harmonically trapped confinement potential along the radial direction.
The triad $({\bf m},{\bf n}, {\bf l})$ yields a particular
texture for each vortex.
Especially, the $l$-vector which is proportional 
to the magnetization gives rise to a texture pattern, which should be
observed by various experimental techniques.

The Bogoliubov equations: eqs.~(\ref{eq:bogo1}) and (\ref{eq:bogo2}) 
which yield the collective excitations have been extended
to the three components.
We calculate the excitations for the above three kinds of the vortices 
and examine the stability condition as a function of the magnetization which 
can be controlled experimentally.
Contrary to the single component BEC, the multi-component BEC allows a
narrow intrinsic stability region for the magnetization, where the multiple components 
work to stabilize the vortex to raise the otherwise negative eigenvalue at 
$q_{\theta}=-1$ to a positive value, that is, 
the non-winding component fills in the core region, effectively serving as a
pinning potential.
This self-organization in the multiple component systems 
for the intrinsic vortex stabilization 
is rather remarkable,
because in the single component BEC the vortex is 
never stabilized without the external forced 
rotation.\cite{isoshima1,isoshima4}
We naively expect that under the forced rotation these vortices enumerated here
are easier to stabilize compare to the single BEC vortex. We will study it near future.





\begin{thebibliography}{1}


\bibitem{firstRb}
M.~H. Anderson, J.~R. Ensher, M.~R. Matthews, C.~E. Wieman and E.~A.~Cornell:
  Science {\bf 269} (1995) 198.

\bibitem{hulet}
C.~C. Bradley, C.~A. Sackett, J.~J. Tollett and R.~G. Hulet:  Phys. Rev. Lett.
  {\bf 75} (1995) 1687.

\bibitem{ketterle}
K.~B. Davis, M.-O. Mewes, M.~R. Andrews, N.~J. van Druten, D.~D. Durfee, D.~M.
  Kurn and W. Ketterle:  Phys. Rev. Lett. {\bf 75} (1995) 3969.


\bibitem{anderson1}
M.~R.~Matthews, B.~P.~Anderson, P.~C.~Haljan, D.~S.~Hall, C.~E.~Wieman and
E.~A.~Cornell: Phys. Rev. Lett. {\bf 83} (1999) 2498.


\bibitem{anderson2}
B.P. Anderson, P.C. Haljan, C.E. Wieman and
E.A.Cornell: Phys. Rev. Lett. {\bf 85} (2000) 2857.


\bibitem{madison1}
K.W. Madison, F. Chevy, W. Wohlleben and J. Dalibard: 
Phys. Rev. Lett. {\bf 84} (2000) 806.


\bibitem{madison2}
K.W. Madison, F. Chevy, W. Wohlleben and J. Dalibard: 
cond-mat/0004037.



\bibitem{madison3}
F. Chevy, K.W. Madison and J. Dalibard: 
Phys. Rev. Lett. {\bf 85} (2000) 2223.


\bibitem{isoshima1}
T. Isoshima and K. Machida: J. Phys. Soc. Jpn. {\bf 66} (1997) 3502.
 
 
 \bibitem{isoshima2}  
T. Isoshima and K. Machida: J. Phys. Soc. Jpn. {\bf 68} (1999) 487.
 
 \bibitem{isoshima3}  
T. Isoshima and K. Machida: Phys. Rev. A {\bf 59} (1999) 2203.

 \bibitem{isoshima4}  
T. Isoshima and K. Machida: Phys. Rev. A {\bf 60} (1999) 3313.

\bibitem{rokhsar}D.S. Rokhsar: Phys. Rev. Lett. {\bf 79} (1997) 2164.
D.~A.~Butts and D.~S.~Rokhsar: Nature {\bf 397} (1999) 327.

\bibitem{dodd}
R. J. Dodd, K. Burnett, M. Edwards and C. W. Clark: Phys. Rev. A {\bf 56} (1997) 587.

\bibitem{doi-natsume}K. Doi and Y. Natsume:
J. Phys. Soc. Jpn. (2001) {\bf 70} in press.

\bibitem{stenger}
J. Stenger, S. Inouye, D.~M. Stamper-Kurn, H.-J. Miesner, A.~P. Chikkatur 
and W. Ketterle: Nature {\bf 369} (1998) 345.
H.-J. Miesner, D.~M. Stamper-Kurn, J. Stenger, S. Inouye,
A.~P. Chikkatur and W. Ketterle:  Phys. Rev. Lett. {\bf 82} (1999) 2228.
D.~M. Stamper-Kurn, H.-J. Miesner, A.~P. Chikkatur, S. Inouye, 
J. Stenger and W. Ketterle:  Phys. Rev. Lett. {\bf 83} (1999) 661.

\bibitem{stamperkurn}
For review of the spinor BEC, see 
D. M. Stamper-Kurn and W. Ketterle: cond-mat/0005001.

\bibitem{ohmi_bec}
T. Ohmi and K. Machida:  J. Phys. Soc. Jpn. {\bf 67} (1998) 1822.

\bibitem{ho}
T.-L. Ho: Phys. Rev. Lett. {\bf 81} (1998) 742.


\bibitem{woelfle}
D. Vollhardt and P. W\"olfle:  {\em The Superfluid Phases of He 3}
  (Taylor-Francis, London, 1990)


\bibitem{yip_vortex}
S.-K. Yip: Phys. Rev. Lett. {\bf 83} (1999) 4677.


\bibitem{aliceString}
U. Leonhardt and G. E. Volovik: JETP Lett. {\bf 72} (2000) 46.



\bibitem{stoof}
H.T.C. Stoof: cond-mat/0002375.


\bibitem{marzlin}
Karl-Peter Marzlin, Weiping Zhang and Barry C. Sanders:
Phys. Rev. A {\bf 62} (2000) 013602


\bibitem{spindomain}
T. Isoshima, K. Machida and T. Ohmi: Phys. Rev. A {\bf 60} (1999) 4857.


\bibitem{isoshimanakahara1}
T. Isoshima, M. Nakahara, T. Ohmi and K. Machida:
Phys. Rev. A {\bf 61} (2000) 063610.


\bibitem{isoshimanakahara2}
M. Nakahara, T. Isoshima, K. Machida, S. Ogawa and T. Ohmi:
Physica B {\bf 284-288} (2000) 17.

\bibitem{double}
T. Isoshima, T. Ohmi and K. Machida:
J. Phys. Soc. Jpn. {\bf 69} (2000) No.12.

\bibitem{martikainen}
J. Martikainen, private communication.

\end{thebibliography}
\end{document}